\documentclass{elsart3p}

\usepackage[square]{natbib}
\usepackage{units}

\usepackage{amssymb}

\usepackage[dvips]{graphicx}
\usepackage{psfrag}
\journal{Nuclear Instruments and Methods in Physics Research, Section A}

\begin{document}

\begin{frontmatter}



\title{Simulations and theory of radio emission from cosmic ray air showers}


\author{T. Huege}
\ead{tim.huege@ik.fzk.de}

\address{Institut f\"ur Kernphysik, Forschungszentrum Karlsruhe, Postfach 3640, 76021 Karlsruhe, Germany}

\begin{abstract}
In the last few years, interest in radio detection of cosmic ray air showers has risen continuously. By now, large-scale application of the radio technique is under investigation in the framework of LOFAR and the Pierre Auger Observatory. The experimental efforts are accompanied by new approaches to describe and model the underlying radiation mechanisms, to lay the foundation for an interpretation of the experimental data. In this article, I review the current radio emission theory and simulations, with slight focus on the geosynchrotron model and its predictions for the information content of the radio signals.
\end{abstract}


\end{frontmatter}

% main text
%________________________________________________________________

\section{Introduction}

Radio detection of cosmic ray air showers promises to nicely complement the existing particle and fluorescence detection techniques for extensive air showers (EAS). It offers 100\% duty cycle, a calorimetric measurement, information about the longitudinal evolution of the shower and very good angular resolution. To exploit these advantages, however, the radio emission physics has to be understood in great detail. Efforts to interpret radio signals from extensive air showers reach back to the early days of radio detection of cosmic rays in the mid-1960s. With the revival of the field in the last few years, interest in simulations and theory of radio emission from EAS has also been increasing continuously. After a very short mention of the historical approaches, I will give an overview of the theory and simulation activities of these last few years and the latest predictions on how to use radio detection to derive air shower parameters of interest.

\section{The early days}

In the year 1962, Askaryan for the first time suggested that charge excess in extensive air showers should produce coherent radio Cherenkov emission in the metre wavelength range. These theoretical predictions \cite{Askaryan1962a,Askaryan1965} led to the first experiments and the successful detection of radio emission from EAS by Jelley et al. \cite{JelleyFruinPorter1965}. After the successful detection, a flurry of activity ensued, both in experimental detection and theoretical studies of the emission mechanism. In 1966, Kahn \& Lerche \cite{KahnLerche1966} proposed that the emission could also be of geomagnetic origin. Their model predicted the emission to arise from the transverse currents of electrons and positrons being deflected in the Earth's magnetic field and the associated dipole propagating through the atmosphere. Many more articles developing models of increasing sophistication were published in the following years (for a review please see \cite{Allan1971}), but the two main lines of thought remained those of Cherenkov radiation and geomagnetic radiation. When the field of radio detection of cosmic rays virtually ceased to exist in the mid-1970s, the consensus, mostly based on polarisation measurements of the emission, was that the radio emission was dominated by a geomagnetic effect.

%(e.g.,\cite{Colgate1967,FujiNishimura1969,CastagnoliSilvestroPicchi1969}

\section{Modern radio theory}

In the year 2001, the experimental efforts for radio detection of cosmic rays with modern digital technology began, and it was clear from the beginning that there was a need for more detailed theory and simulations of radio emission from EAS. Based on the historical results, the expectation was that the radio emission was dominated by a geomagnetic effect.

\subsection{Frequency-domain analytical geosynchrotron model}

\begin{figure}
   \psfrag{nu0MHz}[c][B]{$\nu$~[MHz]}
   \psfrag{Eomega0muVpmpMHz}[c][t]{$\left|\vec{E}(\vec{R},\omega)\right|$~[$\mu$V~m$^{-1}$~MHz$^{-1}$]}
   \psfrag{Enu0muVpmpMHz}[c][b]{$\epsilon_{\nu}$~[$\mu$V~m$^{-1}$~MHz$^{-1}$]}
   \includegraphics[width=\columnwidth]{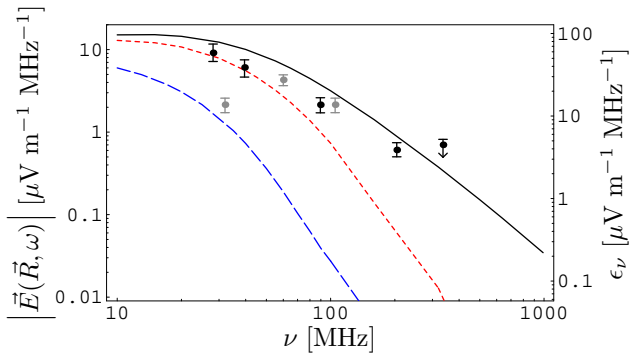}
   \caption{
   \label{fig:spectra_data}
   Electric field frequency spectrum of a vertical $10^{17}$~eV air shower calculated with the analytical geosynchrotron model \cite{HuegeFalcke2003a} in comparison with various historical data. Solid: centre of illuminated area, short-dashed: 100~m from centre, long-dashed: 250~m.
   }
   \end{figure}

\begin{figure}
   \psfrag{R0m}[c][B]{distance from shower centre [m]}
   \psfrag{Eomega0muVpmpMHz}[c][t]{$\left|\vec{E}(\vec{R},\omega)\right|$~[$\mu$V~m$^{-1}$~MHz$^{-1}$]}
   \psfrag{Enu0muVpmpMHz}[c][b]{$\epsilon_{\nu}$~[$\mu$V~m$^{-1}$~MHz$^{-1}$]}
   \includegraphics[width=\columnwidth]{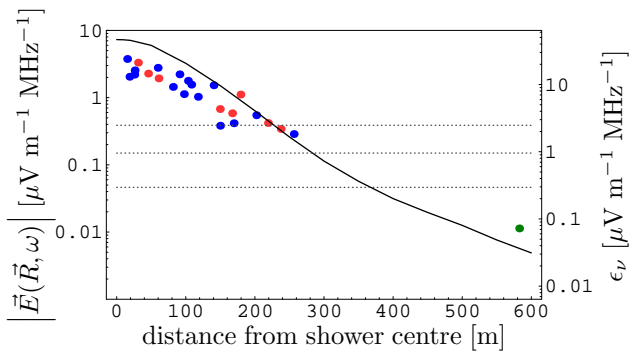}
   \caption{Radial dependence of the 55~MHz electric field component for a vertical $10^{17}$~eV air shower calculated with the analytical geosynchrotron model \cite{HuegeFalcke2003a} in comparison with historical data. Horizontal lines denote predicted detection thresholds for (from top to bottom) one antenna, an array of 10 antennas and an array of 100 antennas.
   \label{fig:radius_dependence_flaring_evolution_integrated_data}}
   \end{figure}

A new approach at the problem of radio emission from cosmic ray air showers was taken with the {\em geosynchrotron model} first proposed in \cite{FalckeGorham2003} and then worked out in detail by Huege \& Falcke in a number of publications \cite{HuegeFalcke2003a,HuegeFalcke2005a,HuegeFalcke2005b}. In addition, the geosynchrotron approach was explored by Suprun, Gorham and Rosner \cite{SuprunGorhamRosner2003}, who however did not follow the topic further. The geosynchrotron model describes the radio emission as coherent, synchrotron-like radiation from the electrons and positrons being deflected in the Earth's magnetic field. In a first step, the emission was calculated with an analytical approach in the frequency domain, based on parameterisations of the relevant air shower properties \cite{HuegeFalcke2003a}. The analytical approach, although naturally simplified, helped understand the important coherence effects present because the scales in the air shower are comparable to the observing wavelengths of the radio emission. A comparison with historical data showed that the analytical model was also able to qualitatively reproduce the frequency (Fig.\ \ref{fig:spectra_data}) and lateral (Fig.\ \ref{fig:radius_dependence_flaring_evolution_integrated_data}) distributions available at that time.

%______________________________________________________________
%______________________________________________________________

%______________________________________________________________
%______________________________________________________________

\subsection{Time-domain Monte Carlo geosynchrotron model (REAS1)}

In a next step, the geosynchrotron model was implemented in a completely different technical way, as a Monte Carlo code (REAS1) working in the time-domain \cite{HuegeFalcke2005a}. Based on the same analytical parameterisations for the underlying air shower properties, the code was able to nicely reproduce the earlier findings of the analytical description. Because experimental data to compare these results to were sparse at that time, this cross-check was an important milestone in the development of the model. In addition, one was now able to simulate air showers also for geometries other than vertical incidence and to produce detailed information on the polarisation of the radio emission. 

\begin{figure}[!ht]
   \centering
   %\psfrag{Eomegaew0muVpmpMHz}[c][t]{$\left|E_{\mathrm{EW}}(\vec{R},2\pi\nu)\right|$~[$\mu$V~m$^{-1}$~MHz$^{-1}$]}   
   \psfrag{Eomegaew0muVpmpMHz}[c][t]{$\left|E(\vec{R},2\pi\nu)\right|$~[$\mu$V~m$^{-1}$~MHz$^{-1}$]}
   \includegraphics[height=\columnwidth,angle=270]{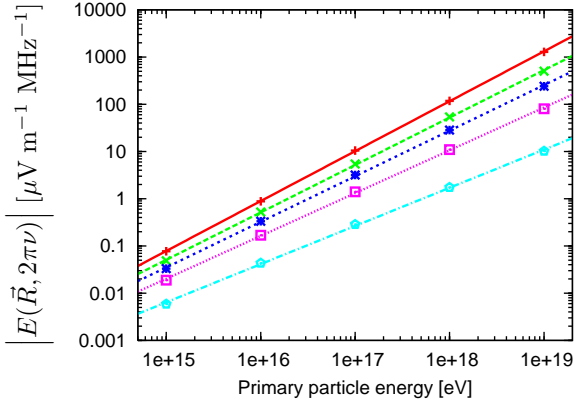}
   \caption[Scaling with primary particle energy]{
   \label{fig:scalingwithep}
   Scaling of the REAS1-simulated 10~MHz electric field emitted by a vertical air shower as a function of primary particle energy $E_{\mathrm{p}}$ \cite{HuegeFalcke2005b}. From top to bottom: 20~m, 100~m, 180~m, 300~m and 500~m to the north from the shower centre.
   }
   \end{figure}

\begin{figure}[!ht]
   \centering
   \includegraphics[height=0.46\columnwidth,angle=270]{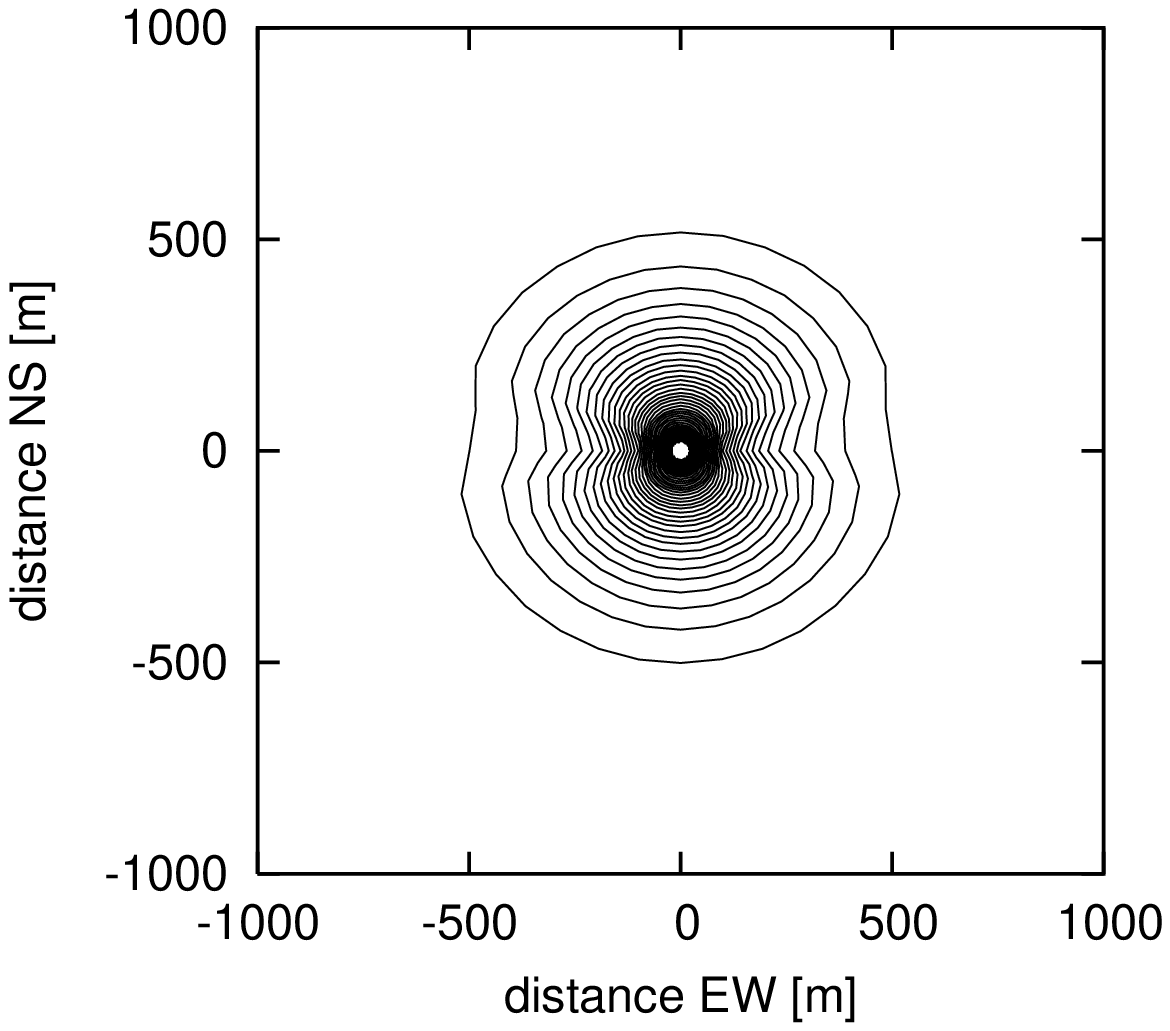}
   \includegraphics[height=0.46\columnwidth,angle=270]{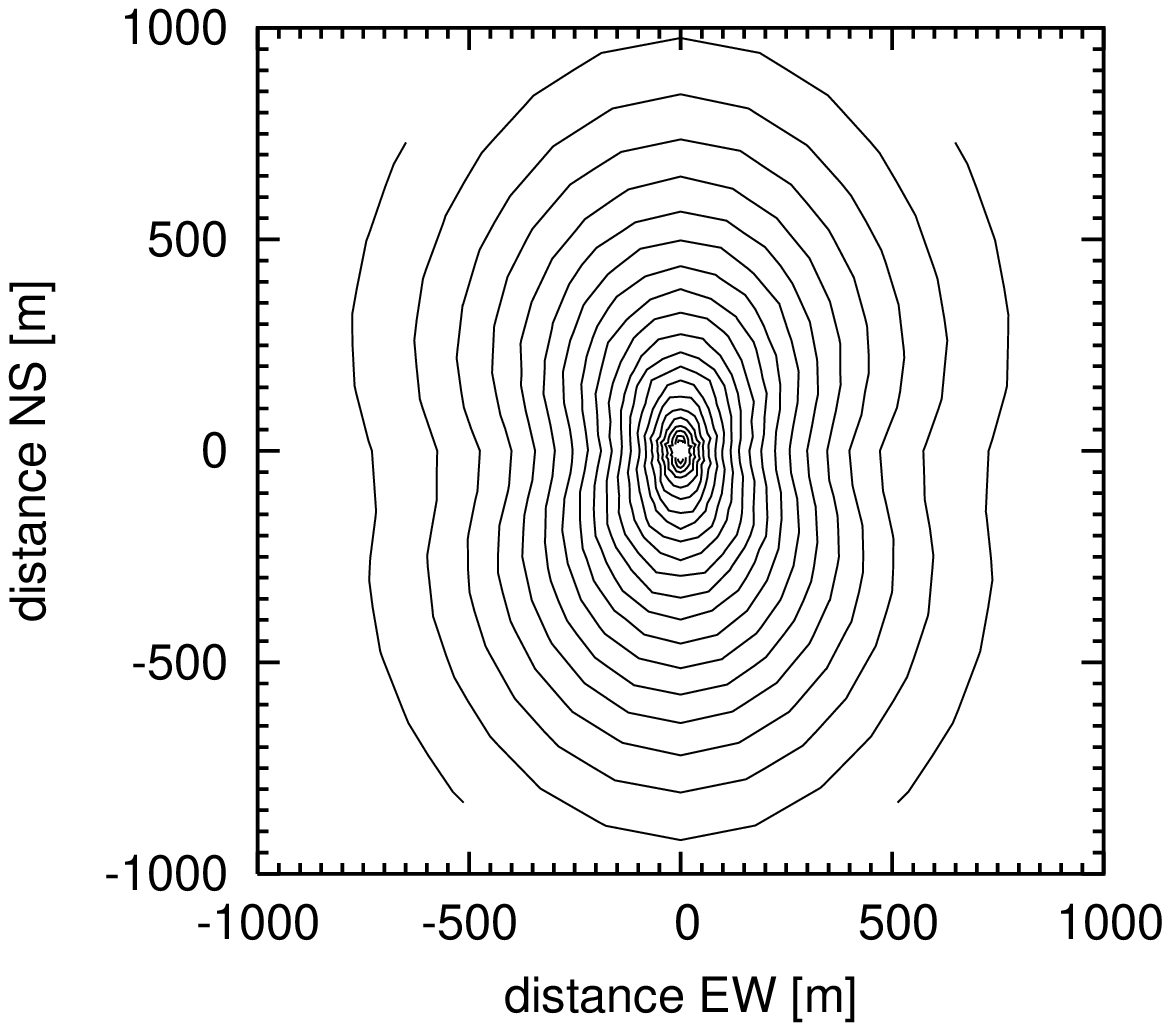}
   \caption[Contour plots of a 10$^{17}$~eV vertical or 45$^{\circ}$ air shower]{
   \label{fig:verticalcontours}
   Contour plots of the 10~MHz absolute field strength for emission from a $10^{17}$~eV vertical air shower (left) and a $10^{17}$~eV shower with 45$^{\circ}$ zenith angle (right) as simulated by REAS1 \cite{HuegeFalcke2005b}. Contour levels are 0.25~$\mu$V~m$^{-1}$~MHz$^{-1}$ apart.
   }
   \end{figure}

\begin{figure}[!ht]
   \psfrag{E0muVpm}[c][t]{$\left|E_{i}(\vec{R},t)\right|$~[$\mu$V~m$^{-1}$]}   
   \psfrag{t0ns}[c][b]{$t$~[ns]}   
   \centering
   \includegraphics[height=\columnwidth,angle=270]{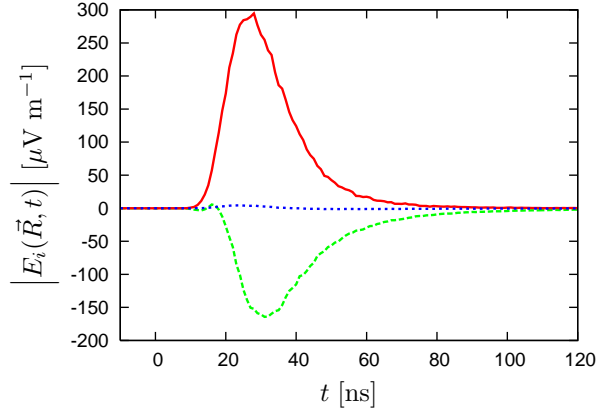}
   \caption[Polarization: raw pulses]{
   \label{fig:pulsespolarisation}
   Raw (unfiltered) pulses in the individual linear polarization components at 200~m distance to the north-west from the centre of a 10$^{17}$~eV vertical air shower simulated with REAS1 \cite{HuegeFalcke2005b}. Solid: east-west component, dashed: north-south component, dotted: vertical component.
   }
   \end{figure}

With the REAS1 code, the predicted radio emission from cosmic ray air showers could for the first time be analysed in detail. Its dependence on important air shower parameters such as shower zenith angle (Fig.\ \ref{fig:verticalcontours}), energy of the primary particle (Fig.\ \ref{fig:scalingwithep}), observing frequency and magnetic field configuration as well as the polarisation characteristics of the radio emission (Fig.\ \ref{fig:pulsespolarisation}) were analyzed and parameterised in a next publication \cite{HuegeFalcke2005b}. In parallel, the prediction that inclined air showers should have a more favourable footprint for radio detection than vertical showers had been independently made by Gousset, Ravel and Roy with an analytical consideration of the geometric effects arising from the air shower geometry \cite{GoussetRavelRoy2004}.

%______________________________________________________________
%______________________________________________________________

%______________________________________________________________
% 4.1 cm for normal style, 3.0 cm for referee style
%______________________________________________________________

%______________________________________________________________
%______________________________________________________________

When the first modern measurements of radio emission from cosmic ray air showers became available at that time from the LOPES \cite{FalckeNature2005} and CODALEMA \cite{ArdouinBelletoileCharrier2005} experiments, interest in the physics of the radio emission rose strongly, and other groups started their own modelling efforts.

\subsection{Frequency-domain EGS-based Monte Carlo}

\begin{figure}[!ht]
   \centering
   \includegraphics[width=\columnwidth]{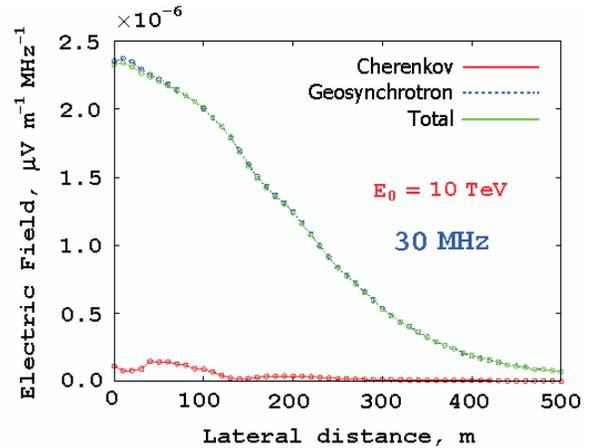}
   \caption{
   \label{kalmykov1}
   Lateral distribution of the 30~MHz radio signal from a 10~TeV vertical air shower simulated with the EGS-based Monte Carlo code of \cite{EngelKalmykovKonstantinovICRC2005}. The geomagnetic component dominates the emission.
   }
   \end{figure}

\begin{figure}[!ht]
   \centering
   \includegraphics[width=\columnwidth]{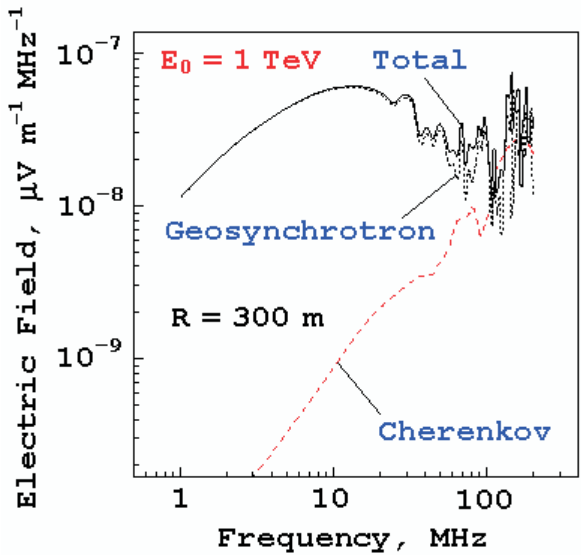}
   \caption{
   \label{kalmykov2}
   Frequency spectra of 1 TeV vertical air showers simulated with the EGS-based Monte Carlo code of \cite{EngelKalmykovKonstantinovICRC2005}.
   }
   \end{figure}

Engel, Kalmykov and Konstantinov tackled the problem with a Monte Carlo approach in the frequency domain \cite{EngelKalmykovKonstantinovICRC2005} based on a special version of the EGS code for the simulation of electromagnetic cascades. While limited by computation time to fairly low energies, their approach allowed a first investigation of the importance of the Askaryan-like Cherenkov contribution, based on a realistic description of the refractive index profile of the atmosphere. Their results confirmed the expectations that the geomagnetic emission dominates over the Cherenkov contributions (Figs.\ \ref{kalmykov1} and \ref{kalmykov2}). Their later analyses showed that it is, however, in fact not straight-forward to cleanly differentiate between a pure geomagnetic and a pure Cherenkov component, because under realistic conditions the two mix and can no longer be clearly disentangled.

%______________________________________________________________
%______________________________________________________________

%______________________________________________________________
%______________________________________________________________

\subsection{Time-domain Monte Carlo geosynchrotron model based on AIRES}

\begin{figure}[!ht]
   \centering
   \includegraphics[width=\columnwidth]{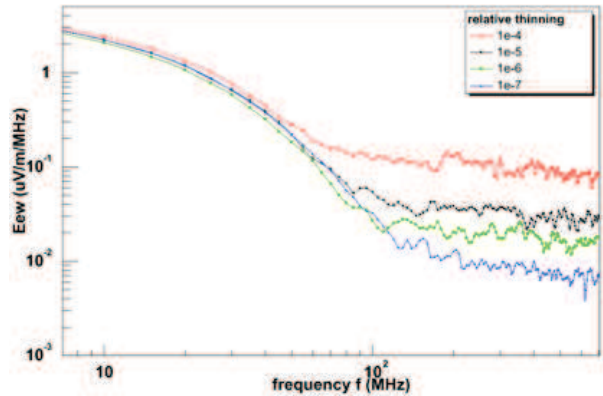}
   \caption{
   \label{duvernois1}
   Radio emission simulated with the AIRES-based code of DuVernois et al. \cite{DuVernoisIcrc2005}, illustrating the influence of particle thinning.
   }
   \end{figure}

DuVernois, Cai and Kleckner took up the concept of time-domain Monte Carlo simulations of geosynchrotron radiation by implementing the effect in the AIRES air shower simulation code \cite{DuVernoisIcrc2005}. The radio emission physics is the same as that performed in the REAS1 code. However, the AIRES-based code uses realistic simulated particle distributions of air showers rather than parameterisations. To make these simulations possible without too large a hit on computation time, thinning techniques are used, the effect of which on the radio emission predictions are visible in Fig. \ref{duvernois1}. The published results \cite{DuVernoisIcrc2005} were qualitatively similar to those of REAS1, although the predicted electric field strengths tended to be about an order of magnitude higher than those calculated with REAS1 (and later REAS2).

%______________________________________________________________
%______________________________________________________________

\subsection{Geosynchrotron spectra of finite particle tracks}

In a paper by Luo \cite{Luo2006}, another look at the basic ingredients of the analytical geosynchrotron model \cite{HuegeFalcke2003a} was taken with an analysis of the effects of finite trajectories on the synchrotron spectrum of particles deflected in the Earth's magnetic field. None of the currently followed approaches are, however, any longer based on these analytical synchrotron spectra.

\subsection{Time-domain Monte Carlo geosynchrotron model based on CORSIKA (REAS2)}

The REAS code made the step to realistic particle distributions based on individual Monte Carlo air shower simulations with the transition from REAS1 to REAS2 performed by Huege, Ulrich and Engel \cite{HuegeUlrichEngel2007a}. In REAS2, the parameterised particle distributions were replaced with realistic distributions simulated with CORSIKA \cite{HeckKnappCapdevielle1998}. Contrary to the AIRES-based code by DuVernois et al., the radio emission routines are, however, not incorporated into CORSIKA directly. Instead, the particle distributions relevant for the calculation of the radio signal are sampled with several multi-dimensional histograms over the course of the air shower simulation with CORSIKA and saved to disk for later usage in the REAS2 code. This approach has the disadvantage that some information is lost during the histogramming process (in particular azimuthal asymmetries in the air shower structure and local over- and underdensities). On the other hand, the approach has several advantages: it allowed the authors to make a very gradual transition from parameterised to histogrammed particle distributions, with a detailed analysis of how this transition changes the radio emission \cite{HuegeUlrichEngel2007a}; powerful optimisation techniques can be applied to cut down the needed computation time by more than a factor of ten; and the radio emission simulations can be studied in detail without the added difficulty of shower-to-shower fluctuations.

\begin{figure*}[!ht]
   \centering
   \includegraphics[width=3.9cm,angle=270]{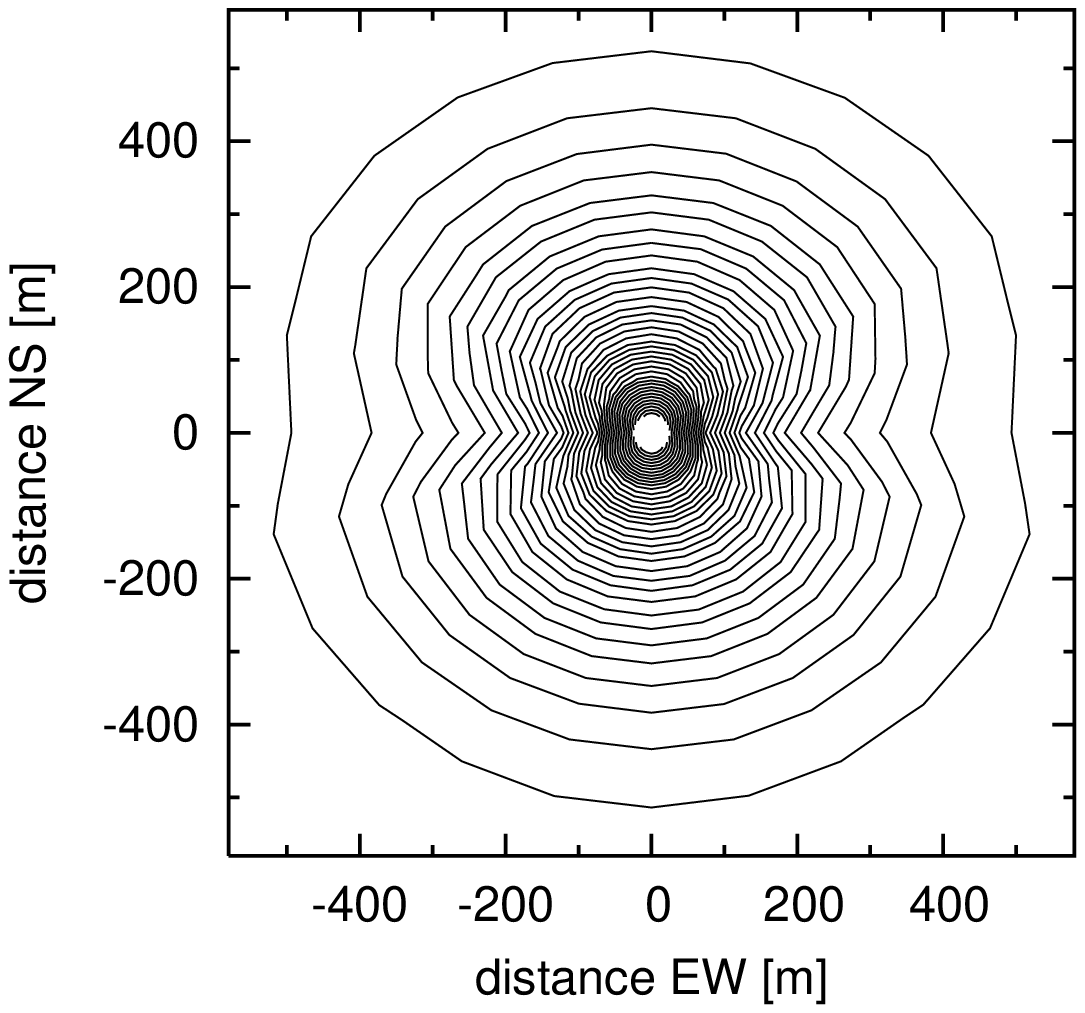}
   \includegraphics[width=3.9cm,angle=270]{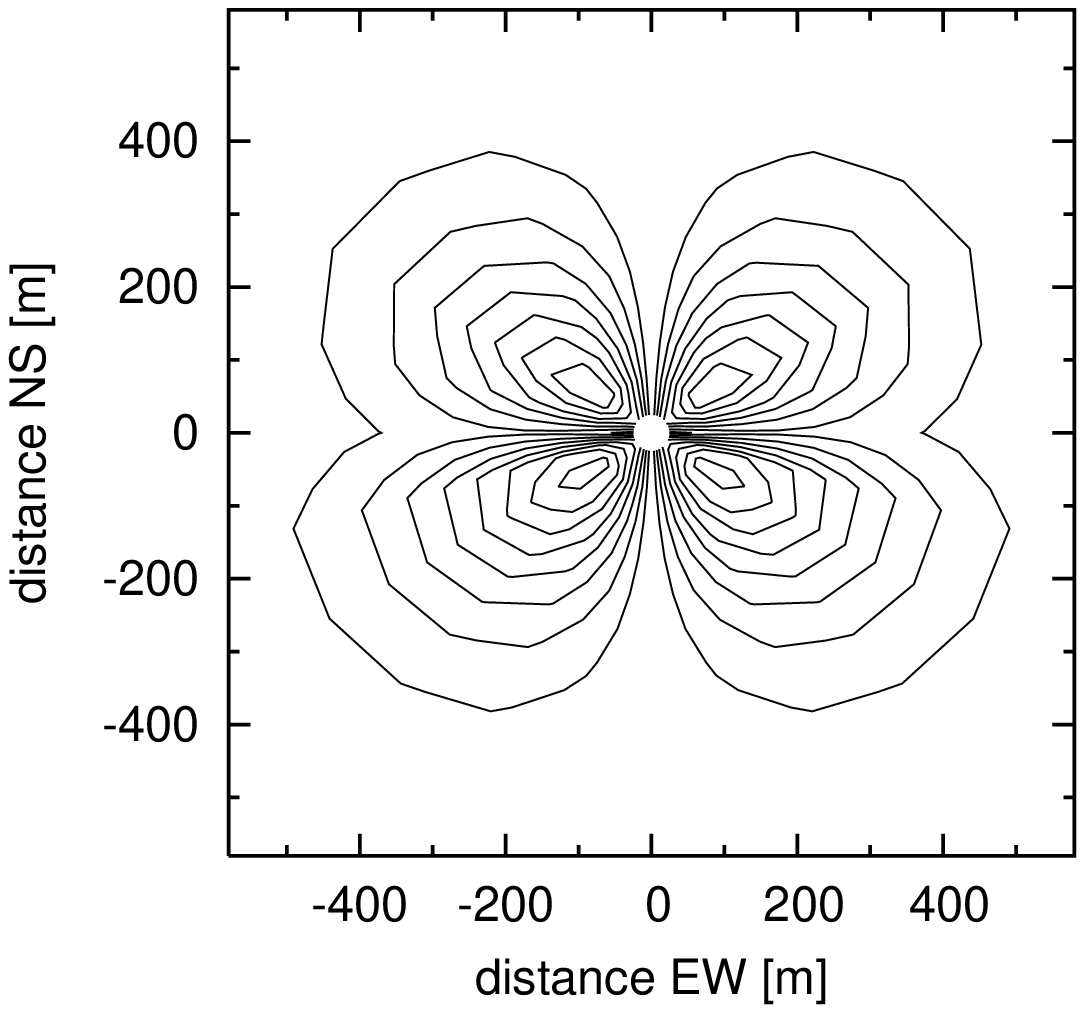}
   \includegraphics[width=3.9cm,angle=270]{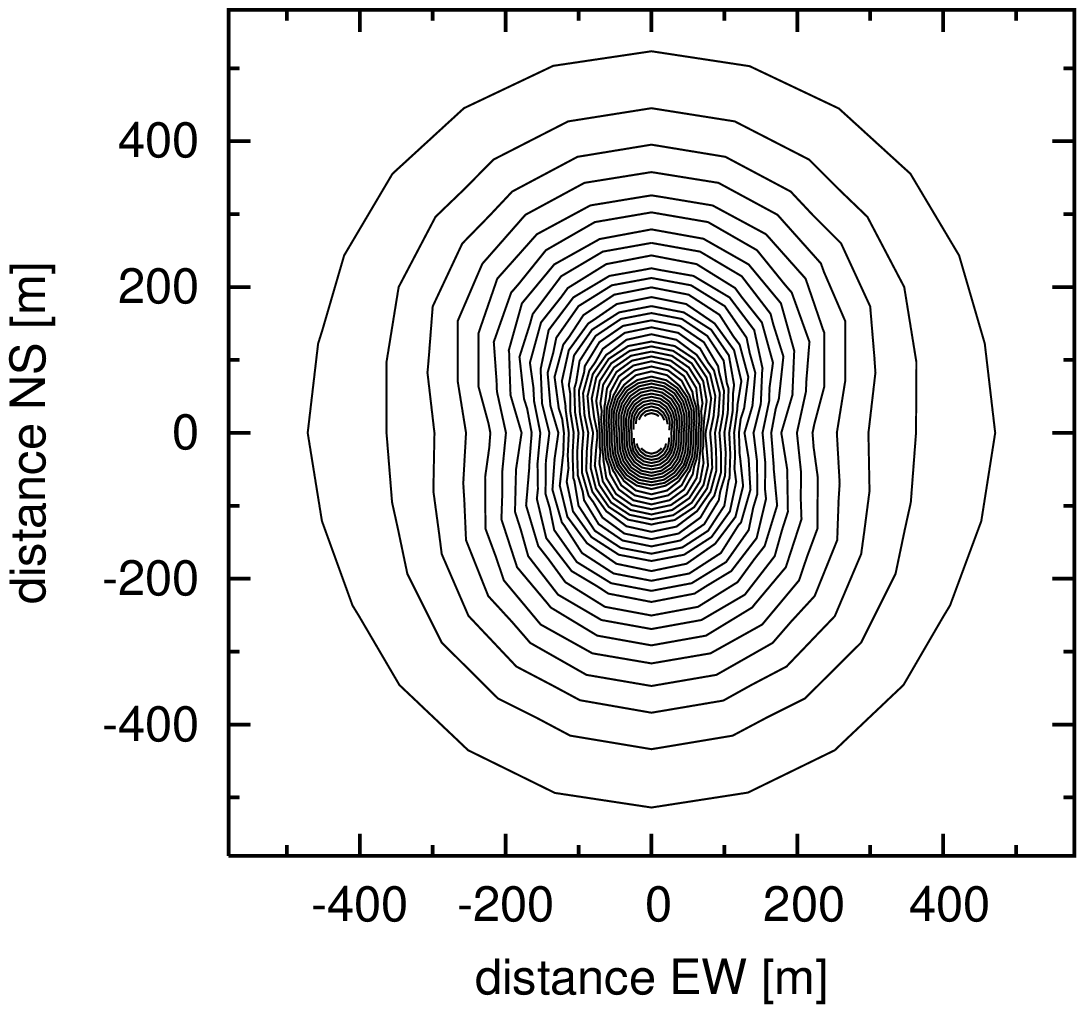}\\
   \includegraphics[width=3.9cm,angle=270]{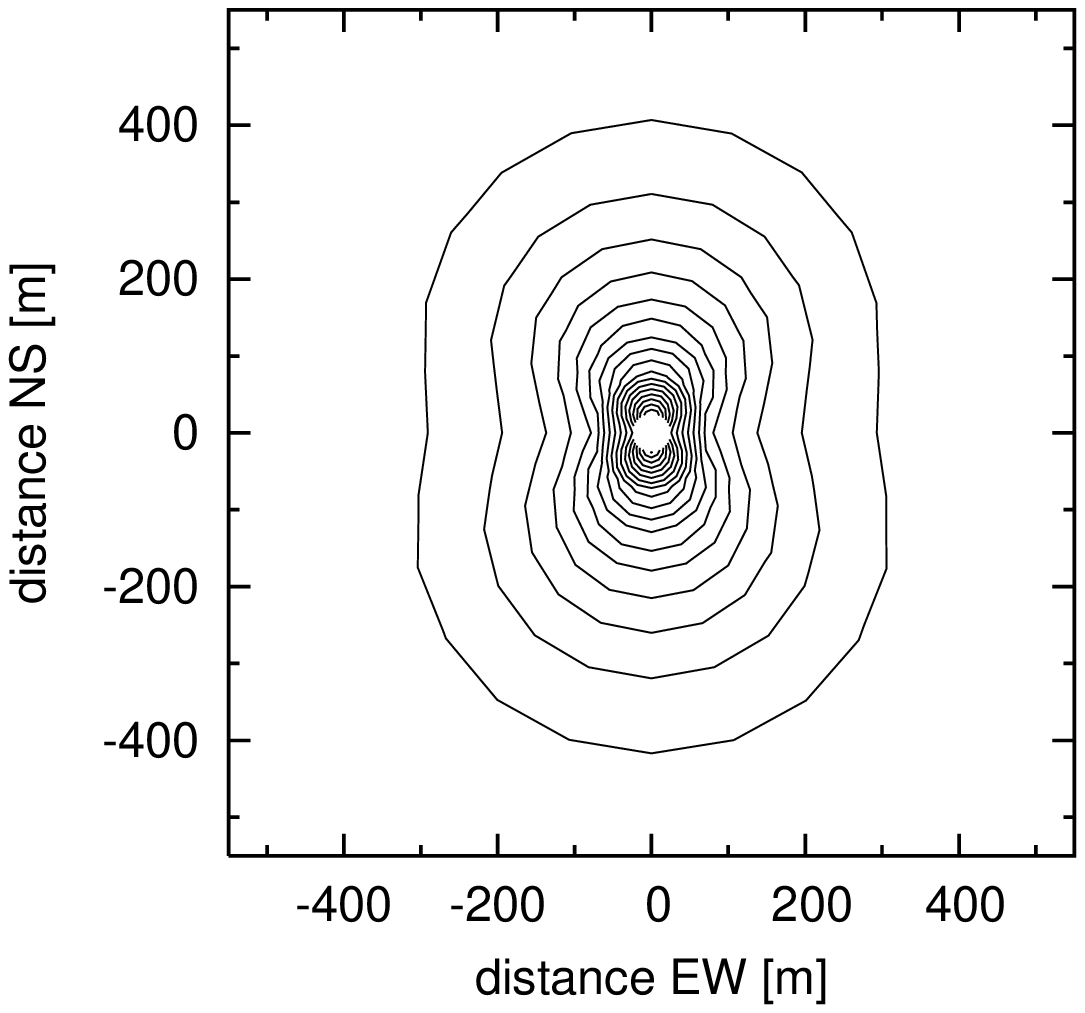}
   \includegraphics[width=3.9cm,angle=270]{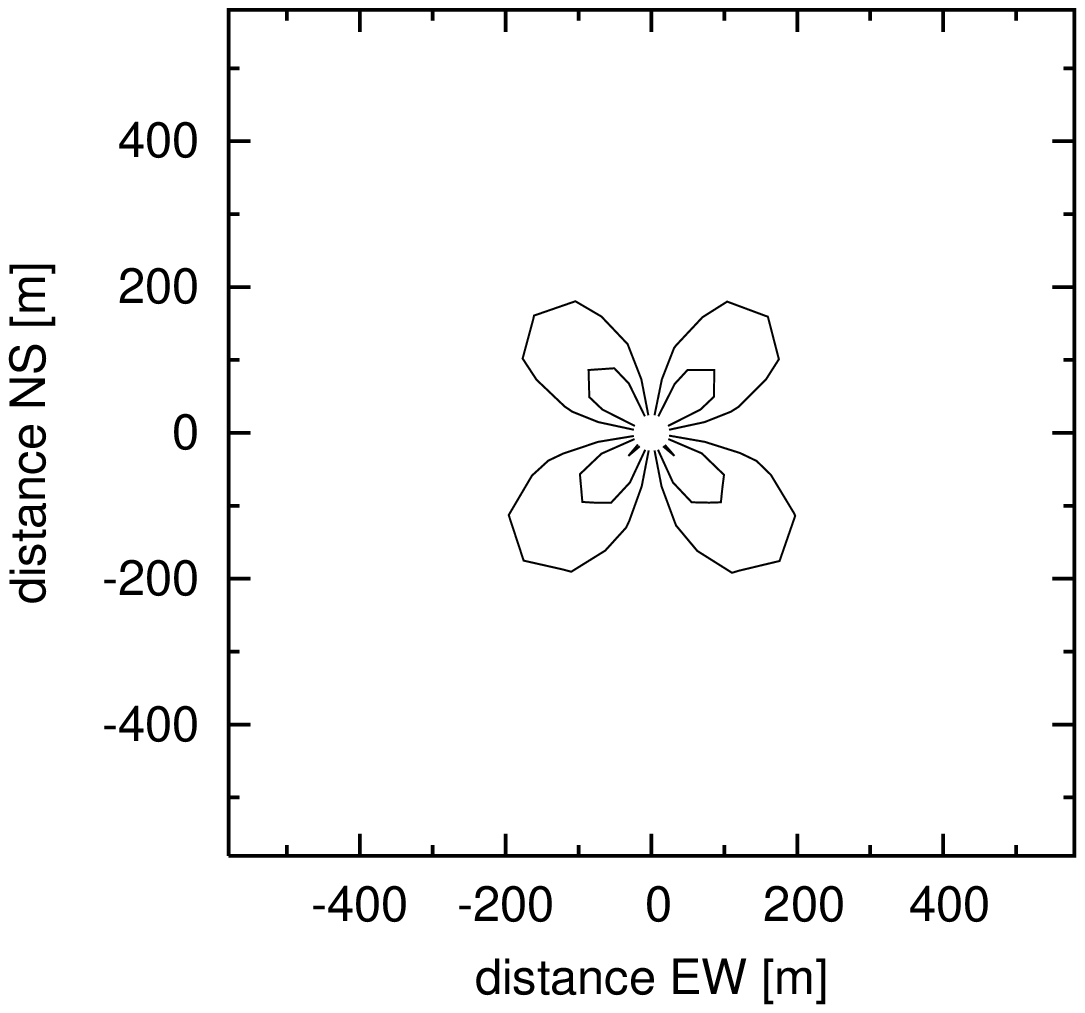}
   \includegraphics[width=3.9cm,angle=270]{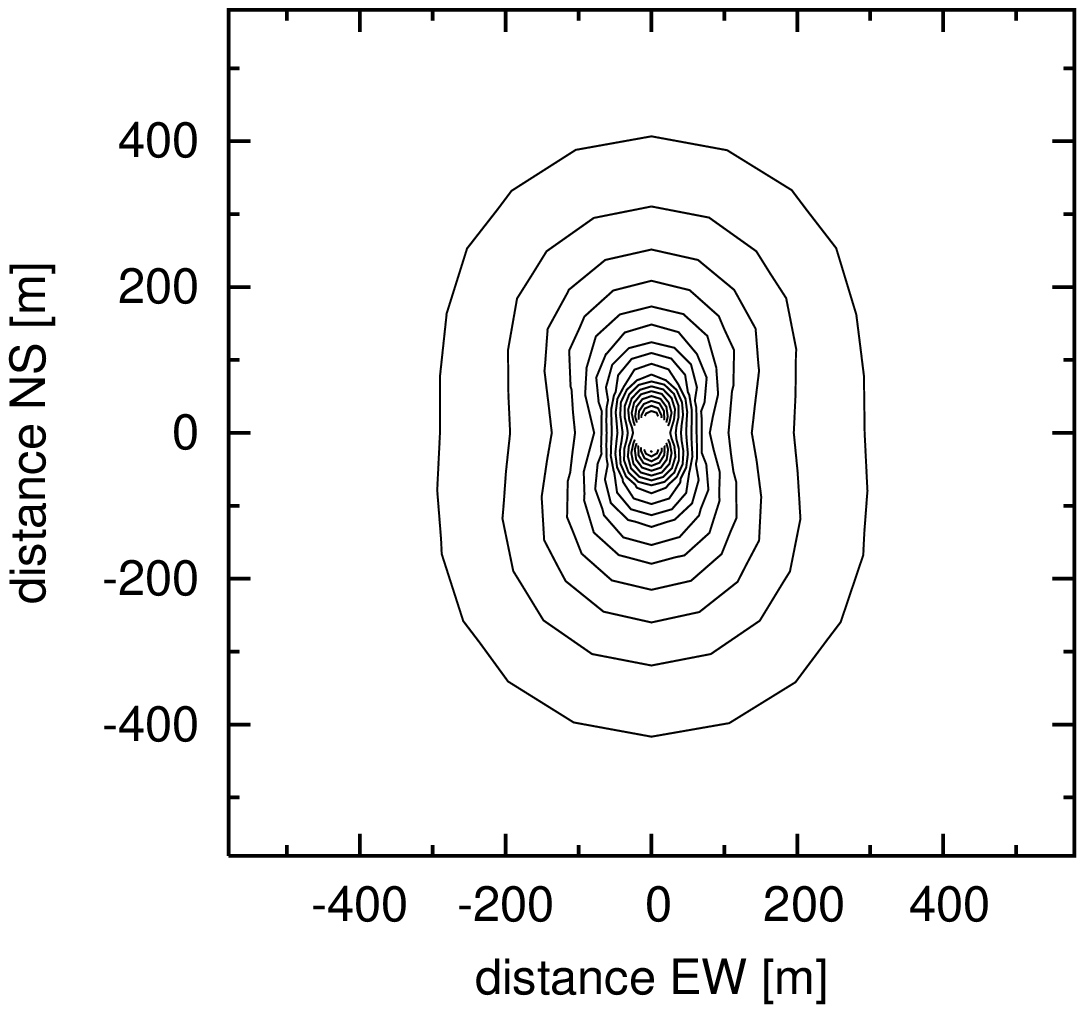}\\
   \caption[Contour plots of the parametrised and histogrammed air shower.]{
   \label{contoursvorhernachher}
   Contour plots of the REAS1-simulated (upper) and REAS2-simulated (lower) air shower emission at $\nu=10$~MHz \cite{HuegeUlrichEngel2007a}. The columns (from left to right) show the total field strength, the north-south polarisation component and the east-west polarisation component. The vertical polarisation component (not shown here) does not contain any significant flux. Contour levels are 0.25~$\mu$V~m$^{-1}$~MHz$^{-1}$ apart, outermost contour corresponds to 0.25~$\mu$V~m$^{-1}$~MHz$^{-1}$.
   }
   \end{figure*}

A comparison of the REAS2 results with those of REAS1 showed that the overall results of REAS2 are similar in the frequency range of interest for radio detection of EAS. The most important differences to REAS1 are a much more pronounced asymmetry of the air shower footprint (Fig.\ \ref{contoursvorhernachher}), shorter pulses at close distances to the air shower core, and a now general prediction of unipolar pulses.

%______________________________________________________________
% 4.1 cm for normal style, 3.0 cm for referee style
%______________________________________________________________

\begin{figure}[htb]
%\begin{minipage}{15.5pc}
%\includegraphics[angle=270,width=15.5pc]{pptdepthregimes75m.eps}
%\caption{\label{depthregimes75m}Contribution of different shower evolution stages to the radio pulse at 75~m north from the shower centre.}
%\end{minipage} \hspace{1.5pc}
%\begin{minipage}{15.5pc}
\centering
\includegraphics[height=\columnwidth,angle=270]{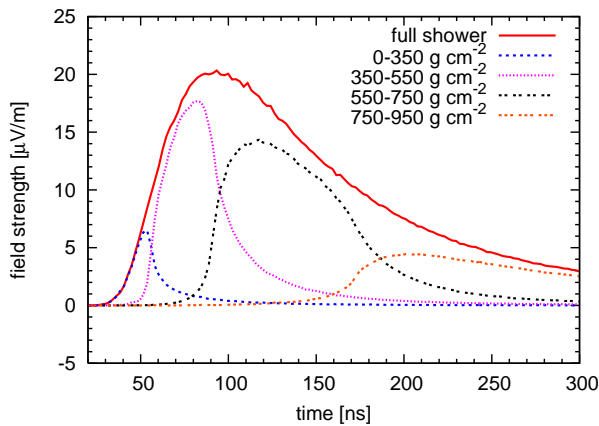}
\caption{\label{depthregimes525m}REAS2-simulated contributions of different shower evolution stages to the radio pulse from a vertical $10^{17}$~eV air shower seen at 525~m north from the shower centre \cite{HuegeUlrichEngel2007a}.}
%\end{minipage}
\end{figure}

The REAS2 code can also be used to study the effects of air shower properties on the radio emission in great detail. One example is shown here in Fig.\ \ref{depthregimes525m}, in which the contributions of different stages of the air shower evolution to the radio pulses at medium lateral distances is illustrated. This demonstrates that information about the air shower evolution is indeed encoded in the time structure of the radio pulses, from which it can in principle be extracted.

\subsection{Boosted Coulomb fields and Cherenkov radiation}

\begin{figure}[!ht]
   \centering
   \includegraphics[width=\columnwidth]{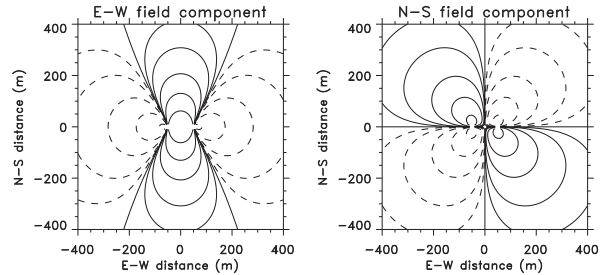}
   \caption{
   \label{meyervernet1}
   Contour plots of the east-west (left) and north-south (right) 20~MHz electric field component of the boosted Coulomb field radio emission from $10^{8}$ electrons and $10^{8}$ positrons as predicted in \cite{MeyerVernetLecacheuxArdouin2008}.
   }
   \end{figure}

In a recent paper \cite{MeyerVernetLecacheuxArdouin2008}, Meyer-Vernet, Lecacheux and Ardouin discuss radio emission from cosmic ray air showers described in the formalism of boosted Coulomb and Cherenkov fields. The authors adopt a simplified air shower model, and the refractive index is approximated as constant over the atmosphere with its value corresponding to that at low to mid altitudes. They conclude that the boosted Coulomb fields predict electric fields of a similar magnitude as those predicted by the geosynchrotron model implemented in REAS. Other properties of the boosted Coulomb fields such as the exponential field strength decrease with lateral distance, the exponential decay of the frequency spectra and the levelling off of the frequency spectra for very low frequencies are in fact also very similar to those of the geosynchrotron model. Another similarity is seen in the polarisation characteristics illustrated with contour plots in Fig.\ \ref{meyervernet1}. The Cherenkov fields are expected by the authors to contribute on a smaller, but not insignificant scale.

%______________________________________________________________
%______________________________________________________________

\subsection{Time-domain macroscopic geomagnetic emission model}

\begin{figure}[!ht]
   \centering
   \includegraphics[width=4.7cm]{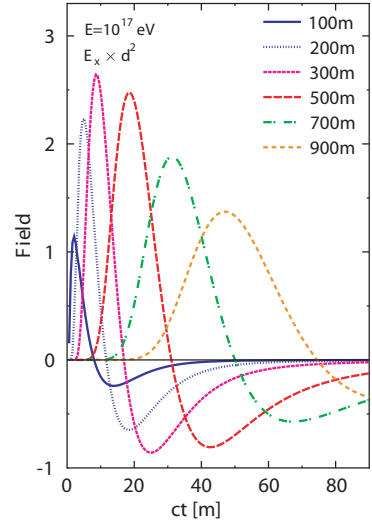}
   \caption{
   \label{scholten1}
   Radio pulses from a vertical $10^{17}$~eV vertical air shower calculated with the macroscopic geomagnetic emission model of Scholten et al. \cite{ScholtenWernerRusydi2008}. Please note that the pulses at different distances have been scaled for clarity.
   }
   \end{figure}

Another approach to describe radio emission from cosmic ray air showers has been taken by Scholten, Werner and Rusydi \cite{ScholtenWernerRusydi2008}. They chose a more macroscopic approach than the geosynchrotron particle-by-particle description by reviving the Kahn \& Lerche \cite{KahnLerche1966} picture of transverse currents arising from the bulk transverse motion of electrons and positrons in the air shower. While they use a fairly simplified air shower model, they demonstrate that in this picture, the time structure of the radio pulses can be related directly to the longitudinal evolution of the air shower, again confirming that this information should be directly extractable from radio measurements (Fig.\ \ref{scholten1}).

One striking feature of the radio signals predicted by their model is that the pulses are generally bipolar. This is a consequence of the fact that their pulses are proportional to the time-derivative of the air shower profile, which is always bipolar. The bipolarity also leads to frequency spectra which decrease towards low frequencies, with the zero-frequency contribution always being exactly zero. The relation of these findings to the contrasting results of the geosynchrotron model are currently under investigation. A reconciliation of these two very different approaches would mark a major breakthrough in the understanding of radio emission from cosmic ray air showers.

\begin{figure*}[!ht]
   \centering
   \includegraphics[width=0.65\textwidth]{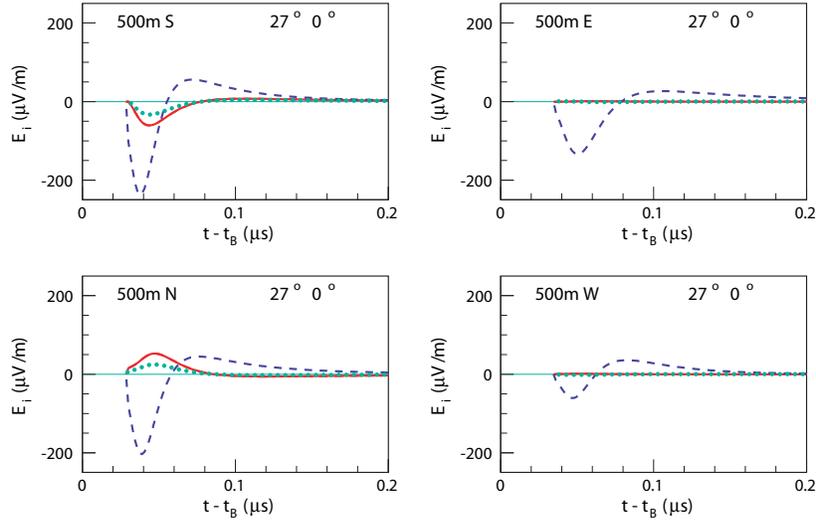}
   \caption{
   \label{werner1}
   Radio pulse polarisation components from a $5 \cdot 10^{17}$~eV air shower seen at 500~m lateral distance to the south, north east and west from the shower core as calculated with the model described in \cite{WernerScholten2008}.
   }
   \end{figure*}

In a second step, Werner \& Scholten have used their approach in combination with a more realistic air shower evolution model, zenith angles other than vertical incidence and a realistic refractive index profile to study the emission physics in more detail. They conclude again that the transverse current  dominates the emission, but that there are corrections which are not negligible \cite{WernerScholten2008} (Fig.\ \ref{werner1}). They state that the effects of the refractive index on the radio emission can be strong, but so far their analysis only applies to point sources without lateral or longitudinal extension.

%______________________________________________________________
%______________________________________________________________

%______________________________________________________________
%______________________________________________________________

\section{Extraction of air shower parameters from radio signals}

With the availability of models incorporating the full complexity of air shower physics, including in particular shower-to-shower fluctuations, it seems worthwhile to investigate the possibilities to extract air shower parameters of interest directly from radio signals. Aside from geometrical parameters such as shower core position and arrival direction, the energy and mass of the primary cosmic ray particle are of particular interest. In a recent paper, Huege, Ulrich \& Engel have used the geosynchrotron model as implemented in the REAS2 Monte Carlo code to investigate this possibility \cite{HuegeUlrichEngel2008}.

\begin{figure}[htb]
%\begin{minipage}{15.5pc}
\centering
\includegraphics[height=\columnwidth,angle=270]{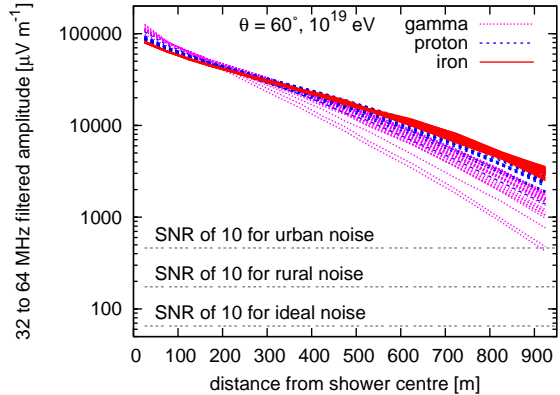}
\caption{\label{lateralslopes}Lateral distribution of the \unit[32--64]{MHz} filtered radio amplitude for \unit[$10^{19}$]{eV} showers coming from the south and observers north of the shower core \cite{HuegeUlrichEngel2008}. Estimates of peak radio amplitudes that would yield a signal-to-noise ratio (SNR) of 10 for ideal (galactic plus atmospheric), rural and urban noise are marked.}
%\end{minipage} \hspace{1.5pc}
%\begin{minipage}{15.5pc}
%\includegraphics[angle=270,width=15.5pc]{60deg0deg_1e19_lateralslopes_maxamp.eps}
%\caption{\label{bla}Bla.}
%\end{minipage}
\end{figure}

\begin{figure}[htb]
\centering
\includegraphics[height=\columnwidth,angle=270]{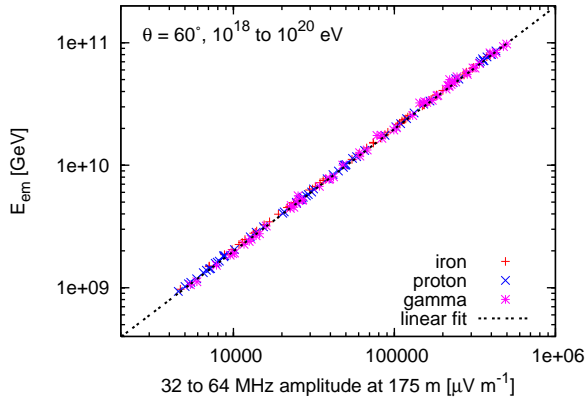}
\caption{\label{energy}A radio measurement in the intersection region directly yields the energy deposited in the atmosphere by the electromagnetic cascade on a shower-to-shower basis \cite{HuegeUlrichEngel2008}. The RMS spread is only $\sim3$\%.}
\end{figure}

They find that the lateral distribution functions of the (bandwidth-filtered) radio signals show a characteristic behaviour. While for each shower the lateral distribution can be described with an exponential function, the slope parameter of this lateral distribution is related to the depth of the underlying air shower's maximum $X_{\mathrm{max}}$. More importantly, there exists a lateral distance region, in which all lateral distribution functions intersect (Fig.\ \ref{lateralslopes}), regardless of primary particle type and shower-to-shower fluctuations. (The location of this region only depends on the shower zenith angle and the radio observing frequency window.) This is even true for air showers with primary particles of different energies, if the field strength profiles are normalised with the energy that the corresponding air shower's electromagnetic cascade deposits in the atmosphere. As a direct consequence, a measurement at this characteristic lateral distance would allow one to directly deduce the energy deposited in the atmosphere by the electromagnetic cascade (Fig.\ \ref{energy}), which in turn can be related to the energy of the primary particle. As shower-to-shower fluctuations do not influence the radio signal in the intersection region, the intrinsic RMS spread of this energy determination is very low ($\sim 3$\%). It should be noted that this is a principle limit, not taking into account any experimental uncertainties.

\begin{figure}[htb]
\centering
\includegraphics[height=\columnwidth,angle=270]{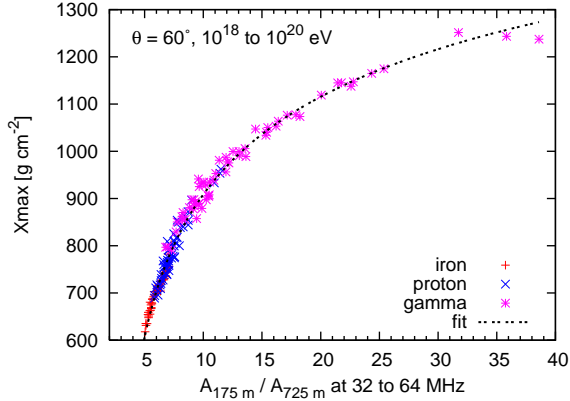}
\caption{\label{composition}The ratio of peak field strengths in the intersection region and at a larger lateral distance yields direct information on the $X_{\mathrm{max}}$ value of an individual air shower \cite{HuegeUlrichEngel2008}.}
\end{figure}

If one measures the slope of the lateral distribution function, e.g., by combining the measurement in the intersection region with one at a larger lateral distance, one can furthermore get a handle on the $X_{\mathrm{max}}$ value of the individual air shower, as shown in Fig.\ \ref{composition}. The intrinsic RMS spread is around 15--20$\,$g$\,$cm$^{-2}$, again not taking into account experimental uncertainties. For comparison, fluorescence measurements yield experimental resolutions of $\sim$35--20$\,$g$\,$cm$^{-2}$. If the value of $X_{\mathrm{max}}$ is known for a specific shower, it can in turn be related to the mass of the primary particle.

While the details (in particular scales) of these predictions do depend on the details of the underlying model, the qualitative behaviour can be considered to be more general, as it is mostly caused by geometrical effects that are not critically dependent on the details of the emission model.

%______________________________________________________________
%______________________________________________________________

%______________________________________________________________
%______________________________________________________________

%______________________________________________________________
%______________________________________________________________

\section{Conclusions}

Radio detection of cosmic ray air showers has once again become a very active field of research, and with it, so have simulations and theory of the radio emission physics. By now, a number of approaches for and technical implementations of the radio emission models exist. In the next few years, the models will have to be compared in detail with the experimental data, which are continuously improving regarding both statistics and quality. At the same time, it is imperative to understand in which way the different models are related to each other and to reconcile contradictions existing between different approaches. In the meantime, the modeling efforts will continue to make very important contributions to our understanding of the radio emission physics and its exploitation for the study of air shower physics.

\begin{ack}
I would like to thank all colleagues contributing to the field, especially those whose results I have used for the preparation of this article.
\end{ack}

% \appendix

% \section{}
% \label{}

%________________________________________________________________________

% Bibliographic references with the natbib package:
% Parenthetical: \citep{Bai92} produces (Bailyn 1992).
% Textual:\cite{Bai95} produces Bailyn et al. (1995).
% An affix and part of a reference:
%   \citep[e.g.][Ch. 2]{Bar76}
%   produces (e.g. Barnes et al. 1976, Ch. 2).

%\bibliography{references}
%\bibliographystyle{elsart-num}

\end{document}